\documentclass[onecoluwn,aps]{revtex4}
\usepackage{graphicx}
\usepackage{amsmath}
\usepackage{amsfonts}
\usepackage{amssymb}
\begin{document}
\draft
\narrowtext
\title{Micellar shape anisotropy and elastic constants in discotic lyotropic liquid crystals}
\author{P. A. de Castro, A. J. Palangana, and L. R. Evangelista}
\address{Departamento de F\'{\i}sica, Universidade Estadual de Maring\'a,\\
Avenida Colombo, 5790, Maring\'a, Paran\'a, Brazil.}

\begin{abstract}
The elastic constants of a discotic lyotropic nematic liquid crystal are
calculated by means of a pseudo-molecular approach as functions of the
micellar shape anisotropy. By assuming that the temperature dependence of
the ratio of the elastic constants comes from the temperature dependence of
the micellar shape anisotropy, the theoretical predictions are connected
with experimental measurements for the ratio $K_{33}/K_{11}$. This procedure
permits to determine, in a phenomenological way, the temperature dependence
for the ratio of elastic constants and for the micellar shape anisotropy
near the nematic-isotropic transition in agreement with the experimental
data.
\end{abstract}

\maketitle
\narrowtext

Lyotropic nematic liquid crystals (NLC) are formed by mixtures of
amphiphilic molecules\cite{de Gennes} and a solvent (usually water), under
convenient temperature and concentration conditions. The basic units of
these systems are anisotropic micelles\cite{Hendrikx}. Under concentration
change, or temperature change, the system can exhibit two uniaxial
and one biaxial phases\cite{Yu}. In the temperature-concentration
phase diagram, the biaxial phase appears to be an intermediate phase between
the uniaxial ones. The uniaxial nematic phases are known as discotic ($
N_{d}$), and calamitic ($N_{c}$), depending on their magnetic anisotropy\cite
{Hendrikx}. For what concerns the elastic properties of a NLC, one of the
fundamental aspects of the research is the determination of the Frank
elastic constants. For a distorted NLC the spatial variations of the
director are small over molecular dimensions and a continuum elastic theory
can be applied. In the bulk of a NLC, in general, can occur three basic
types of distortions: splay, twist, and bend. To these distortions are
associated, respectively, the elastic constants $K_{11}$, $K_{22}$ and $
K_{33}$. These elastic constants have been the object of extensive
experimental and theoretical studies specially when dealing with
thermotropic NLC. An important parameter in this study, in compounds with a
rigid molecular structure, is related to the molecular length-to-width ratio
\cite{Chandrasekhar}.

Experimental results concerning lyotropic systems\cite{Charvolin} show that
the shape anisotropy of the micelles varies according to the composition and
temperature. The micellar shape anisotropy ($R$) is defined as the ratio
between the micellar diameter and the bilayer thickness. Its variation in a
discotic nematic phase as a function of the temperature was determined in
order to investigate a similar behavior of the orientational order parameter
\cite{Galerne}. In this direction, it can be very instructive to investigate
the dependence of the elastic constants on the anisotropy in the micellar
shape of a lyotropic medium. In the literature \cite{Pinto}, very few
investigations have been reported on the elastic constants associated to the
lyotropic systems. In particular, Haven, Armitage and Saupe\cite{Haven} have
carried out precise measurements on the ratio of the elastic constants $
K_{33}$ and $K_{11}$ as a function of the temperature for the discotic
mixture of the decylammonium chloride $(DACl)$, obtaining values comparable
to those determined in nematic thermotropics.

>From the theoretical point of view, it has been recently utilized a
molecular approach to evaluate the micellar shape anisotropy dependence of
the bulk elastic constants in a discotic nematic, by using the Maier-Saupe
interaction law\cite{Evangelista,Maier-Saupe}. In that approach, the value
for the ratio of the elastic constants $K_{33}/K_{11}$ was determined as an
increasing function of the parameter $R$. The tendency and the order of
magnitude of the ratio of elastic constants was found to be in a reasonable
qualitative agreement with some available experimental results and with some
calculations for thermotropics\cite{Govers}. However, in that calculation
the value for this ratio was limited to a saturation value $K_{33}/K_{11}=2$
, thus preventing a broad comparison with other experimental data.

In this paper we apply the molecular approach to calculate the elastic
constants as a function of the micellar shape anisotropy having in mind a
lyotropic nematic phase whose micelles are in the shape of a disk. We
consider a special kind of interaction law which is a mixing between the
Maier-Saupe and Nehring-Saupe interaction laws\cite{Sub,Hat}. For this
interaction law the ratios between the elastic constants are not limited to
these relatively small values for lyotropic NLC. Furthermore, we use the
experimental data of ref.~\cite{Haven} to obtain a phenomenological
dependence of the ratio of the elastic constants as a function of the
temperature, by assuming that this dependence comes from the temperature
dependence of the micellar shape anisotropy.

To determine the elastic constants of a discotic lyotropic we employ the
pseudo-molecular approach, which is an approximate technique to determine
the macroscopic properties of the system from the intermolecular interaction
giving rise to the nematic phase\cite{Nehring,Barberi,Faetti}. In this
approximated method the elastic energy density can be written as

\begin{equation}  \label{hat-8}
f= f_0 + L_{ik} n_{i,k} + N_{ijk} n_{i,jk} + M_{ijkm} n_{i,k} n_{j,m},
\end{equation}
where
\begin{equation}  \label{hat-9}
L_{ik} = \frac{1}{2} \int_{v^{\prime}} q_i u_k r dv^{\prime},
\end{equation}

\begin{equation}  \label{hat-10}
N_{ijk} = \frac{1}{4} \int_{v^{\prime}} q_i u_j u_k r^2 dv^{\prime},
\end{equation}
and
\begin{equation}  \label{hat-11}
M_{ijkm} = \frac{1}{4} \int_{v^{\prime}} q_{ij} u_k u_m r^2 dv^{\prime},
\end{equation}
are elastic tensors. In the above expressions,
\begin{equation}  \label{hat-3}
q_i= \left( \frac{ \partial g} {\partial n_i^{\prime}}\right) _{\vec{n}
^{\prime}=\vec{n}} \quad {\rm and} \quad q_{ij} = \left( \frac{ \partial^2 g
}{ \partial n_i^{\prime}\partial n_j^{\prime}} \right) _{\vec{n} ^{\prime}=
\vec{n}}.
\end{equation}
The derivatives of the interaction energy $g (\vec{n}, \vec{n}^{\prime}, \vec{r})$ are
evaluated on the reference state and the summation convention is assumed. Furthermore,
$\vec{u}=\vec{r}/r$, and hence $x_{k}=u_{k}r$.

The integrations are performed over the interaction volume. Since the
micelles we are considering are disk-like, we consider in the calculations
an interaction volume in the shape of a disk. In assuming an interaction
volume of this form we suppose that the interaction energy $g(\vec{n},\vec{n}
^{\prime },\vec{r})$ is different from zero in the region limited by two
similar disks, whose inner part coincides with the micellar volume, and the
outer part is defined by the long range part of the intermolecular
interaction. For simplicity, the two disks are supposed to be similar,
having the same shape anisotropy, $R=2a/d=2A/D$, where $a$ and $d$ refer to
the inner (micellar) volume, whereas $A$ and $D$ refer to the outer volume.
The bilayer thickness $d$ is supposed to remain essentially independent of
the temperature, according to the experimental results\cite{Galerne}.

We will consider an intermolecular interaction in the mixed form\cite
{Sub,Hat}

\begin{equation}  \label{hat-16}
g(\vec{n}, \vec{n}^{\prime}, \vec{r} ) = -\frac{C}{r^6} [\vec{ n}\cdot\vec{n}
^{\prime}- 3 \epsilon (\vec{n}\cdot\vec{u}) (\vec{n} ^{\prime}\cdot \vec{u}
)]^2,
\end{equation}
where we have introduced a mixing parameter $\epsilon $ such that for $ \epsilon=0$ the
interaction is reduced to the Maier-Saupe law\cite {Maier-Saupe}

\begin{equation}
g_{{\rm MS}}=-\frac{C}{r^{6}}(\vec{n}\cdot \vec{n}^{\prime })^{2},
\label{hat-14}
\end{equation}
whereas for $\epsilon =1$ it is reduced to the Nehring-Saupe law\cite {Nehring}
\begin{equation}
g_{{\rm NS}}=-\frac{C}{r^{6}}[\vec{n}\cdot \vec{n}^{\prime }-3(\vec{u}\cdot
\vec{u}^{\prime })(\vec{u}\cdot \vec{n})]^{2}.	\label{hat-15}
\end{equation}
In the above expressions we have assumed perfect nematic order, i.e., we assume that the
scalar order parameter $S=1$.

The details of the calculations of the elastic constants as functions of the
eccentricity of the micellar (molecular) volume can be found in ref.~\cite
{Hat}. To obtain the elastic constants as a function of the micellar shape
anisotropy one has to perform the integrations over an interaction volume of
discotic shape, characterized by a shape anisotropy $R$. To perform the
integrations in order to evaluate the elements of the elastic tensors we
choose the $z$ axis of the coordinate system along $\vec{n}$. In cylindrical
coordinates $x= \rho \cos\theta$, $y=\rho \sin\theta$, and $z$, we have

\begin{equation}  \label{mic1}
u_1 = \frac{x}{r} = \frac{\rho \cos\theta}{r}, \quad u_2 = \frac{y}{r} =
\frac{\rho \sin\theta}{r}, \quad u_3 = \frac{z}{r},
\end{equation}
and, hence $N = \vec{n}\cdot \vec{u} = z/r$, where $r = \sqrt{\rho^2+z^2}$. In the
calculations we need to evaluate integrals of the type~\cite{Hat}

\begin{equation}  \label{mic2}
I = \int_{v^{\prime}} f(\rho,\theta,z)\rho d\rho d\theta dz,
\end{equation}
with $f(\rho,\theta,z)$ a continuous function of the variables. It is
convenient to put the integrations in the form

\begin{eqnarray}  \label{mic3}
I &=& \int_{v^{\prime}} f(\rho,\theta,z) \rho d\rho dz	\nonumber \\
&=&\int_0^{2\pi} d\theta \left\{ \int_{-D/2}^{D/2} dz \int_0^A
f(\rho,\theta,z) \rho d\rho - \int_{-d/2}^{d/2} dz \int_0^a f(\rho,\theta,z)
\rho d\rho \right\},
\end{eqnarray}
which is simply the integral over the outer volume minus the integral over the inner
(micellar) volume of the same functions of ref.~\cite{Hat}, now in a cylindrical
geometry. The integrations are easily performed and yield the following ratio for the
elastic constants
\begin{equation}  \label{mic5}
\frac{K_{33}} {K_{11}} = \frac{p_{31}}{ q_{31}},
\end{equation}
where
\begin{eqnarray}  \label{mic5a}
p_{31} &=& 4 (R (32 (1 + R^2)^2 (1 + 2 R^2) - 48 \epsilon (1 + 8 R^2 + 11
R^4 + 4 R^6)  \nonumber \\
&+& 3 \epsilon^2 (3 + 136 R^2 + 141 R^4 + 48 R^6)) + (32 - 48 \epsilon + 9
\epsilon^2) (1 + R^2)^3 \arctan(1/R))  \nonumber \\
q_{31} &=& (R (64 (1 + R^2)^2 (3 + 2 R^2) - 128 \epsilon (3 + 8 R^2 + 9 R^4
+ 4 R^6)  \nonumber \\
&+& 9 \epsilon^2 (21 + 56 R^2 + 123 R^4 + 48 R^6)) + 3 (64 - 128 \epsilon +
63 \epsilon^2) (1 + R^2)^3 \arctan(1/R)).
\end{eqnarray}

Let us consider first the case $\epsilon = 0$, which corresponds to the
Maier-Saupe interaction. In this limit we obtain

\begin{equation}
\frac{K_{33}}{K_{11}} = \frac{2(R+2R^3 + (1+R^2) \arctan(1/R)} {R(3+2R^2) +
3(1+R^2) \arctan(1/R)}.
\end{equation}
In the other limit, for $\epsilon = 1$, which corresponds to the
Nehring-Saupe interaction, we obtain

\begin{equation}
\frac{K_{33}}{K_{11}}=\frac{4(7R-152R^{3}-55R^{5}-16R^{7}+
(7+21R^{2}+21R^{4}+7R^{6}) \arctan(1/R)} {
3R+8R^{3}-403R^{5}-48R^{7}+(3+9R^{2}+9R^{4}+3R^{6}) \arctan(1/R)}.
\end{equation}
\begin{figure}[htbp!]
\begin{center}
\includegraphics[width=9cm,angle=-90]{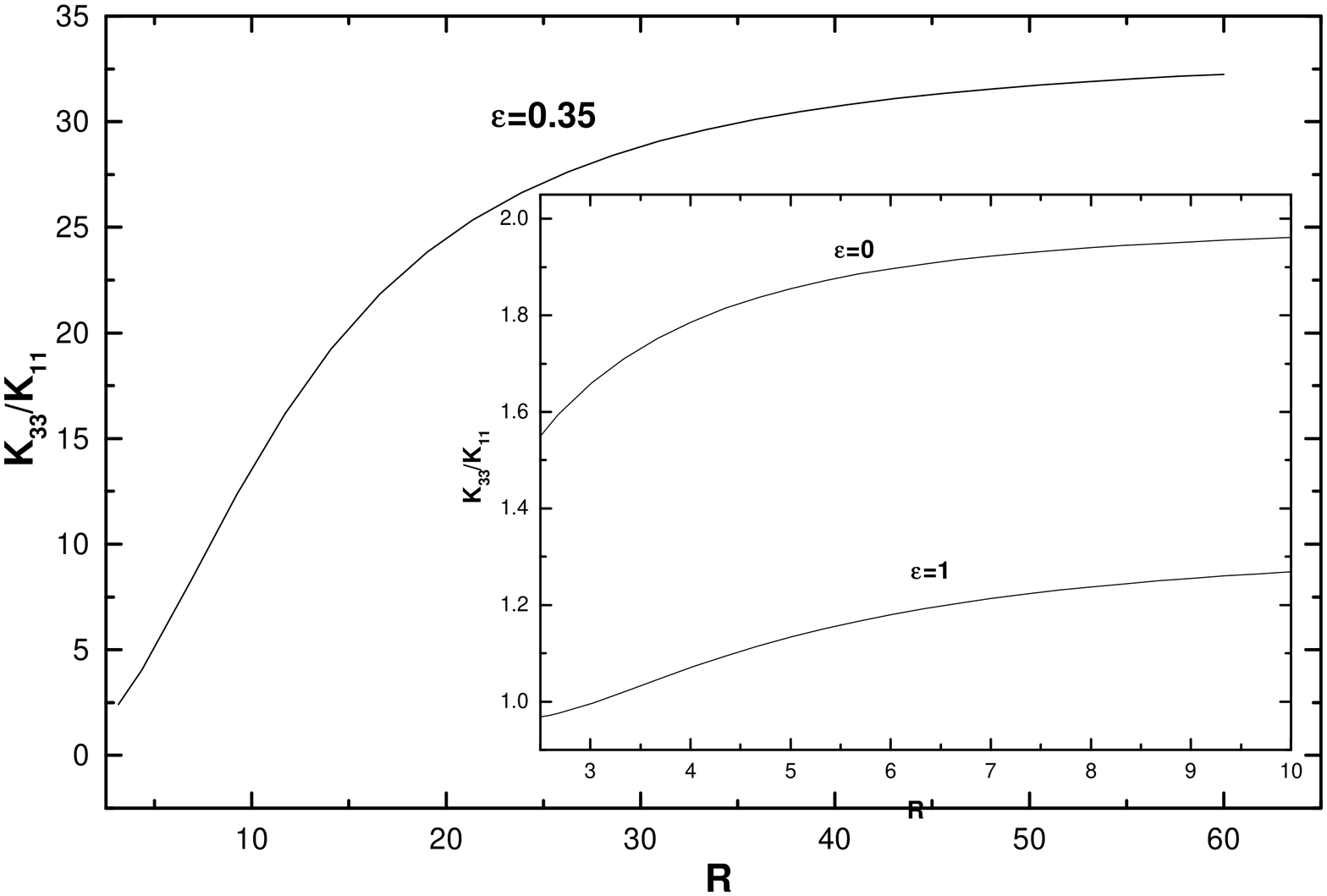}
\end{center}
\caption{The ratio of bend and splay elastic constant vs. the micellar shaper anisotropy
$R$ for distinct values of the mixing parameter $\epsilon$ (see
Eq.~(\protect\ref{hat-16})).} \label{fig1}
\end{figure}
These results indicate that the ratio of the elastic constants is an increasing function
of the micellar shape anisotropy\cite{Evangelista} but it tends to a saturation in both
cases (see Fig.~\ref{fig1}). In fact, when we consider $ R\rightarrow \infty $ we obtain

\begin{equation}  \label{mic7}
\frac{K_{33}}{K_{11}}=\frac{4(2-3\epsilon )^{2}}{8-32\epsilon +27\epsilon^{2}
},
\end{equation}
and the saturation value obviously depends on the mixing parameter $\epsilon$ . For
$\epsilon = 0$, $K_{33}/K_{11} = 2$, whereas for $\epsilon = 1$, $ KK_{33}/K_{11} = 4/3$.
Of course very high values for $R$ make no physical sense. However, as we will discuss
below, in order to account for the experimental values of the ratio $K_{33}/K_{11}$ in a
discotic phase, the saturation value obtained in the above calculation cannot be
restricted to the interval $4/3 \leq K_{33}/K_{11} \leq 2$. To account for these
experimental data we consider the results for the interaction law having a mixing
parameter $\epsilon = 0.35$. For this value, the predicted saturation value for the ratio
$K_{33}/K_{11} \approx 33.6$. It is important to stress again that the ``pure''
Maier-Saupe interaction and the ``pure'' Nehring-Saupe interaction cannot furnish a ratio
of this magnitude, since, as stressed before, they have saturation values which are
small, when compared with the ones experimentally found.

Our procedure to obtain the trend with temperature for the ratio $
K_{33}/K_{11}$ is the following. We assume that the dependence of this ratio
with the micellar shape anisotropy is given by Eqs.~(\ref{mic5}) and (\ref
{mic5a}). From the experimental data of ref.~\cite{Haven} we obtain the
micellar shape anisotropy $R$ for the corresponding value of the reduced
temperature $T/T_{\rm{NI}}$, where $T_{\rm{NI}}$ is the nematic-isotropic transition
temperature. This result in turn furnishes the trend of $K_{33}/K_{11}$
versus the reduced temperature as predicted by the theoretical calculation
we have performed.

In \cite{Haven} the splay and bend elastic constants have been measured, as a
function of the temperature, for the mixture $DACl$, NH$_{4}$Cl, and water,
in a nematic discotic phase. From these experimental data it is possible to
obtain the ratio $K_{33}/K_{11}$, studied by the authors, as a function of
the temperature for two compositions: (mixture 1) $DACl$ 7.00, NH$_{4}$Cl
2.53, H$_{2}$O 90.40 mole \%; this sample presents the following sequence of
phases: neat soap, 35 $^{\circ}$C, nematic, 55 $^{\circ}$C, isotropic;
(mixture 2) $DACl$ 7.57, NH$_{4}$Cl 2.73, H$_{2}$O 89.70 mole \%; neat soap,
45.5 $^{\circ}$C, nematic, 65.5 $^{\circ}$C, isotropic.
Their results shown that, in the vicinity of the
neat soap phase the bend elastic constant diverges while the splay constant
shows no critical behavior. Among the measured
values for the ratio, the higher one is $K_{33}/K_{11} \approx 5.3$ for
mixture 1 and $K_{33}/K_{11} \approx 19.3$ for mixture 2.

\begin{figure}[htbp!]
\begin{center}
\includegraphics[width=9cm,angle=-90]{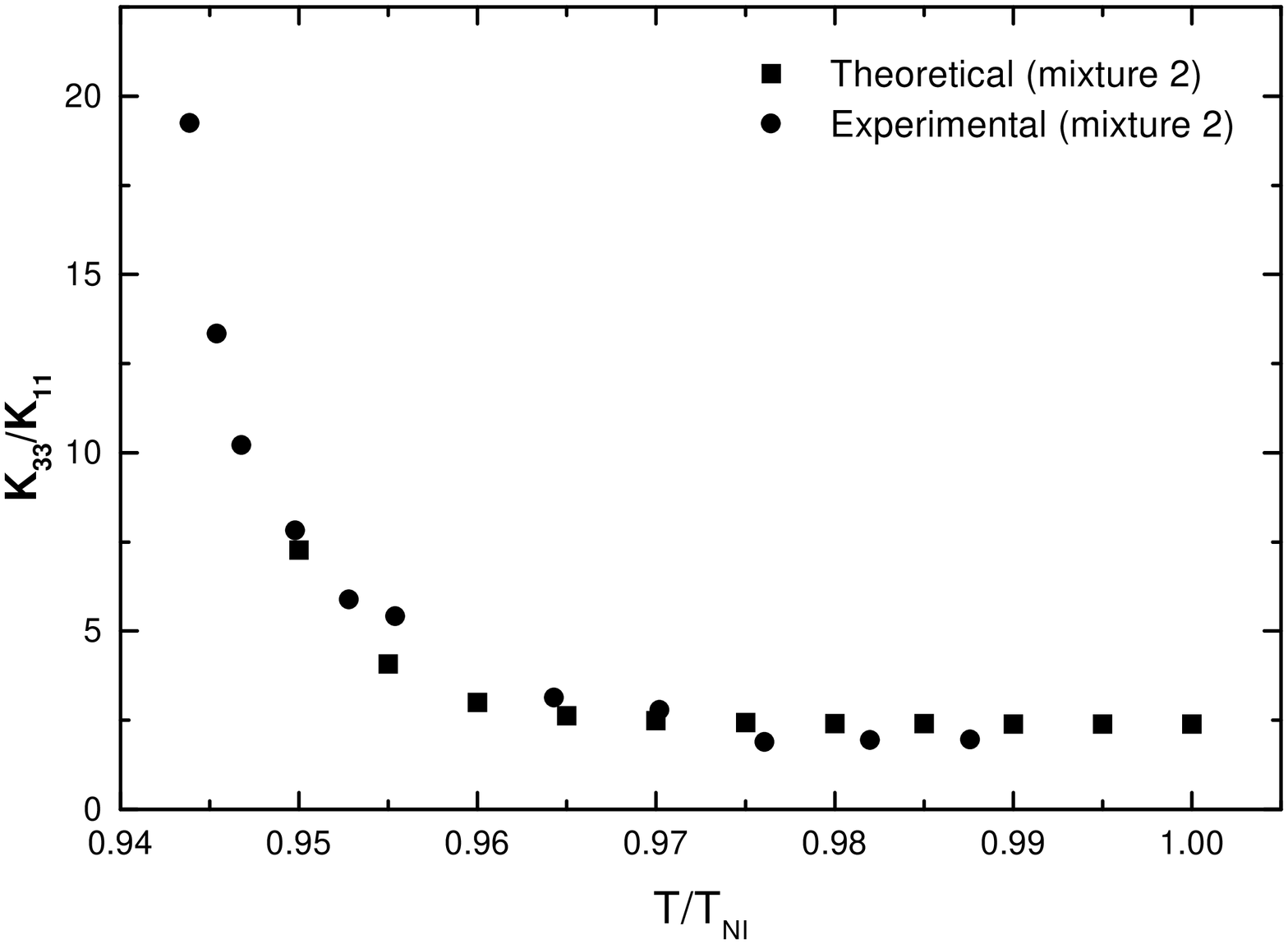}
\end{center}
\caption{The ratio of bend and splay elastic constant vs. temperature for mixture 1 as
predicted by this work ($\times$). The experimental data ($\triangle$) comes from
ref.\protect\cite{Haven}. $T_{\rm{NI}}$ is the nematic-isotropic transition temperature}
\label{fig2}
\end{figure}

The ratio $K_{33}/K_{11}$ versus the reduced temperature of the theoretical
calculations (for a mixing parameter $\epsilon =0.35$) and the experimental
data for mixture 1 are shown in Fig.~\ref{fig2}. The ratios, both
experimental and theoretical, fall on the same curve. A similar behavior has
been observed also for mixture 2. With this procedure we have obtained, in a
phenomenological way, the dependence of the micellar shape anisotropy $R$ on
the reduced temperature for a discotic nematic phase. Furthermore, X-ray
diffraction results \cite{Galerne} performed in a nematic discotic lyotropic
mixture (potassium laurate - KL, 1-decanol-DeOH and D$_{2}$O - mixture 3)
show that the temperature variations of the shape anisotropy and of the
orientational order parameter are similar within the experimental error. The
bilayer thickness is found to remain independent of temperature and in this
lyotropic system the nematic discotic exhibits a reentrant behavior.

\begin{figure}[htbp!]
\begin{center}
\includegraphics[width=9cm,angle=-90]{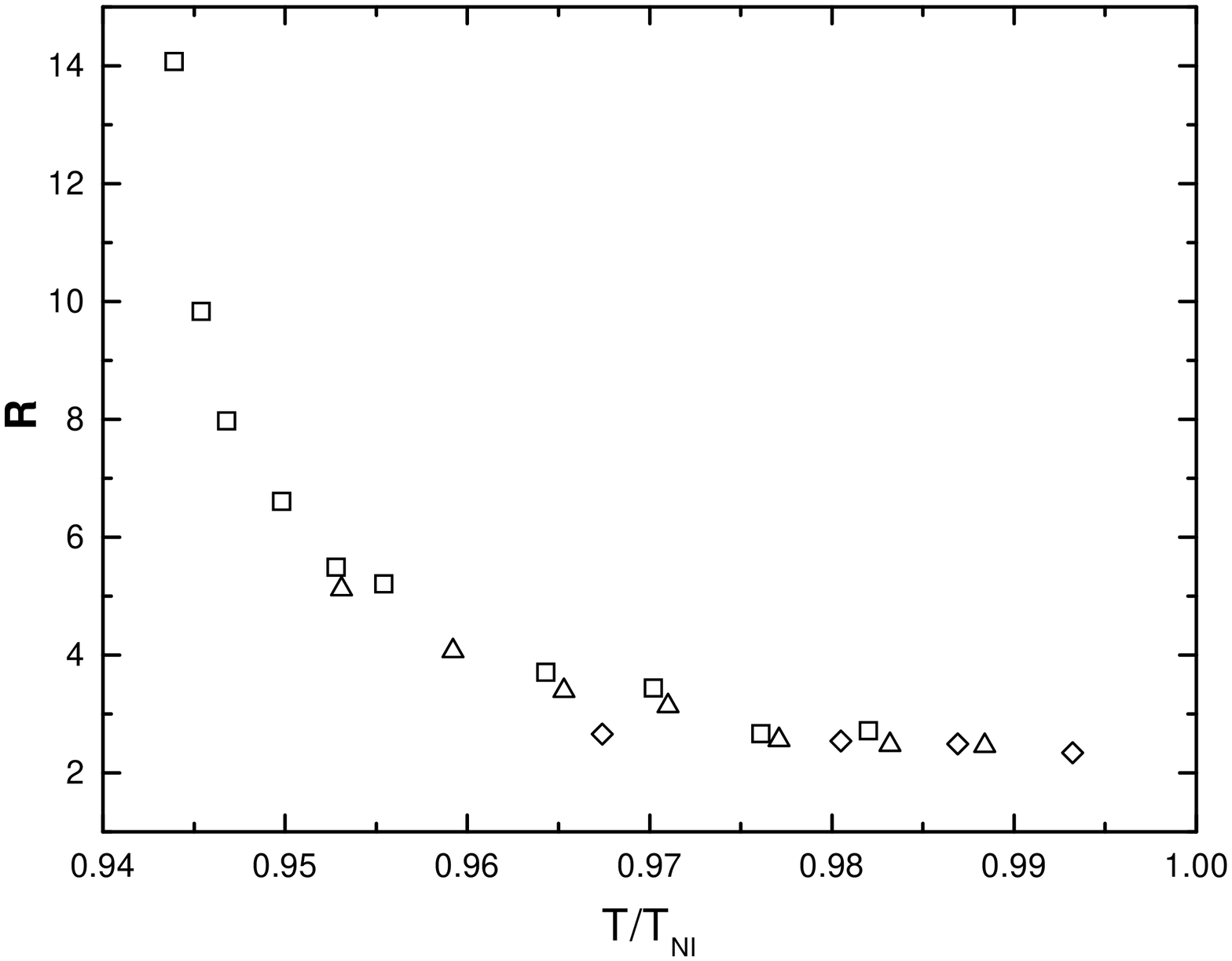}
\end{center}
\caption{The trend of the micellar shape anisotropy $R$ vs. temperature. The curves for
mixtures 1 ($\triangle$) and 2 ($\Box$) comes from our theoretical predicitions (best
fits) whereas the curve for mixture 3 ($\Diamond$) is the experimental results from
ref.\protect\cite{Galerne}.} \label{fig3}
\end{figure}

The connection between experimental and theoretical results for the ratios $R$ versus
reduced temperature are presented in Fig.~\ref{fig3}. Note that, on coming from the
discotic nematic to the isotropic phase our results agree well with earlier measurements
of the shape anisotropy on the mixture 3, particularly in vicinity of nematic-isotropic
phase transition. A more precise comparison between our theoretical calculations with the
experimental data is difficult since there are a few data available on the measurements
of the shape anisotropy in lyotropic systems. The results show another interesting
physical behavior since the temperature dependence of the ratio comes from the micellar
shape anisotropy dependence on the temperature instead of the scalar order parameter
dependence. In fact, the temperature dependence of the bulk elastic constants comes
(approximated) from its quadratic dependence on the scalar order parameter. Therefore,
this dependence is not decisive in the ratio of two bulk elastic constants of
thermotropic NLC (it is particularly weak near the nematic-isotropic transition
\cite{deJeu}). But the experimental data reported above for lyotropic NLC shows that this
dependence exists and is also very strong. In this manner, it seems to be meaningful to
consider that the temperature dependence for lyotropic NLC can be due mainly to the
micellar shape anisotropy behavior with the temperature. Another important point to be
considered here is the introduction of the mixing parameter $\epsilon $ on the
intermolecular interaction law among micelles of the lyotropic medium which seems to be
the key of this study. The choice of $\epsilon =0.35$ is just an indication that ``pure''
laws like the Maier-Saupe and Nehring-Saupe ones can not account for the experimental
results, and  the mixed interaction has to be taken into account.

\acknowledgments

This work has been partially supported by the Brazilian Agency, CNPq.

\end{document}